\documentclass[aps,prl,twocolumn,superscriptaddress,showkeys]{revtex4-1}

\usepackage{color}
\usepackage{graphicx}
\usepackage{dcolumn}
\usepackage{bm}
\usepackage[english]{babel}
\usepackage[latin1]{inputenc}
\usepackage[bottom=2cm,top=2cm, left=1.5cm, right=1.5cm]{geometry}
\usepackage{booktabs}
\usepackage{amsmath}
\usepackage[unicode=true, bookmarks=true, bookmarksnumbered=false, bookmarksopen=false, breaklinks=false, pdfborder={0 0 1}, backref=false, colorlinks=false]{hyperref}

\begin{document}

\title{Fermi Volume Evolution and Crystal-Field Excitations 
in Heavy-Fermion Compounds Probed by Time-Domain Terahertz Spectroscopy}

\author{S. Pal}
\affiliation{Department of Materials, ETH Z\"urich, 8093 Z\"urich, Switzerland}

\author{C. Wetli}
\affiliation{Department of Materials, ETH Z\"urich, 8093 Z\"urich, Switzerland}

\author{F. Zamani}
\affiliation{Physikalisches Institut and Bethe Center for Theoretical Physics, 
Universit\"{a}t Bonn, Nussallee 12, 53115 Bonn, Germany}

\author{O. Stockert}
\affiliation{Max Planck Institute for Chemical Physics of Solids, 01187 Dresden, Germany}

\author{H. v. L\"{o}hneysen}
\affiliation{Institut f\"{u}r Festk\"{o}rperphysik and Physikalisches Institut, Karlsruhe Institute of Technology, 76021 Karlsruhe, Germany}

\author{M. Fiebig}
\email{manfred.fiebig@mat.ethz.ch}
\affiliation{Department of Materials, ETH Z\"urich, 8093 Z\"urich, Switzerland}

\author{J. Kroha}
\email{kroha@physik.uni-bonn.de}
\affiliation{Physikalisches Institut and Bethe Center for Theoretical Physics, Universit\"{a}t Bonn, Nussallee 12, 53115 Bonn, Germany}
\affiliation{Center for Correlated Matter, Zhejiang University, Hangzhou, Zhejiang 310058, China}

\date{March 7, 2019}

\begin{abstract}
We measure the quasiparticle
weight in the heavy-fermion compound CeCu$_{6-x}$Au$_{x}$ ($x=0,\ 0.1$) by
time-resolved terahertz spectroscopy for temperatures from 2 up to 300\,K. 
This method distinguishes contributions from the heavy Kondo band and from the
crystal-electric-field satellite bands by different terahertz response delay
times. We find that the formation of heavy bands is controlled by an
exponentially enhanced, high-energy Kondo scale once the 
crystal-electric-field states become
thermally occupied. We corroborate these observations by temperature-dependent 
dynamical mean-field calculations for the multiorbital Anderson lattice model 
and discuss consequences for quantum-critical scenarios. 
\end{abstract}


\maketitle

In heavy-fermion materials \cite{Loehneysen07}, a lattice of rare-earth ions
with local magnetic moments in the $4f$ shell is embedded in a metallic host. 
With decreasing temperature, the Kondo effect drives part of the $4f$ spectral 
weight to the Kondo resonance near the Fermi energy $\varepsilon_{F}$ 
\cite{Hewson93} where it forms a band of lattice-coherent, heavy
quasiparticles (QPs). Consequently, part of the $4f$ electrons become itinerant,
and the Fermi volume expands so as to accommodate the extra number of
indistinguishable $4f$ electrons in the Fermi sea. The existence of an enlarged
Fermi volume is, therefore, a unique signature of the Kondo-induced heavy
Fermi-liquid phase, its absence a signature of heavy-QP destruction
\cite{Benlagra11,Choi12}, as it may occur, e.g., near a quantum phase
transition (QPT) \cite{Coleman01,Hackl08,Friedemann10}. Within the standard
Anderson lattice model, the crossover energy scale above which the Kondo
correlations fade away is the Kondo lattice temperature $T_K^*$. However, the
recent observation of a large Fermi surface in the heavy-fermion compound
YbRh$_2$Si$_2$ at temperatures $T\gg T_K^*$ by angle-resolved photoemission
spectroscopy (ARPES) \cite{Kummer15} has raised disputes about this
picture \cite{Choi12,Kummer15,Paschen16,Feng17}. 
In particular, it has been questioned whether $T_K^*$, as extracted from 
low-temperature thermodynamic and transport measurements, 
is the correct scale for heavy-QP formation, 
or whether they can persist to much higher energies \cite{Kummer15}. 
This question is important, because the behavior of the QP formation scale near a heavy-fermion QPT is a hallmark distinguishing different quantum-critical scenarios, like the spin-density-wave scenario \cite{Hertz76,Moriya85,Millis93}, the local quantum-critical scenario \cite{Coleman01,Si01}, or other schemes of QP destruction \cite{Senthil04,Woelfle11,Nejati17}.

In this Letter, we resolve this puzzle by separately measuring the Kondo and
the crystal-electric-field (CEF) contributions to the Fermi volume using 
time-resolved terahertz spectroscopy \cite{Wetli18} and 
temperature-dependent dynamical mean-field theory (DMFT) calculations. 
Time-domain terahertz spectroscopy has been recently developed as a method 
particularly sensitive to the  QP dynamics in strongly correlated electron
systems \cite{Wetli18}. 
We find that for the heavy-fermion compound CeCu$_{6-x}$Au$_x$, the spectral
weight contributing to the large Fermi volume at high temperatures is
accounted for by the CEF satellite resonances of the Ce $4f$ orbitals, while the
low-temperature behavior is controlled by the Kondo resonance, in particular
near the QPT at $x=0.1$. This reconciles the seemingly
contradictory ARPES results \cite{Kummer15}.

\begin{figure}[t!]
\includegraphics[width=0.9\columnwidth]{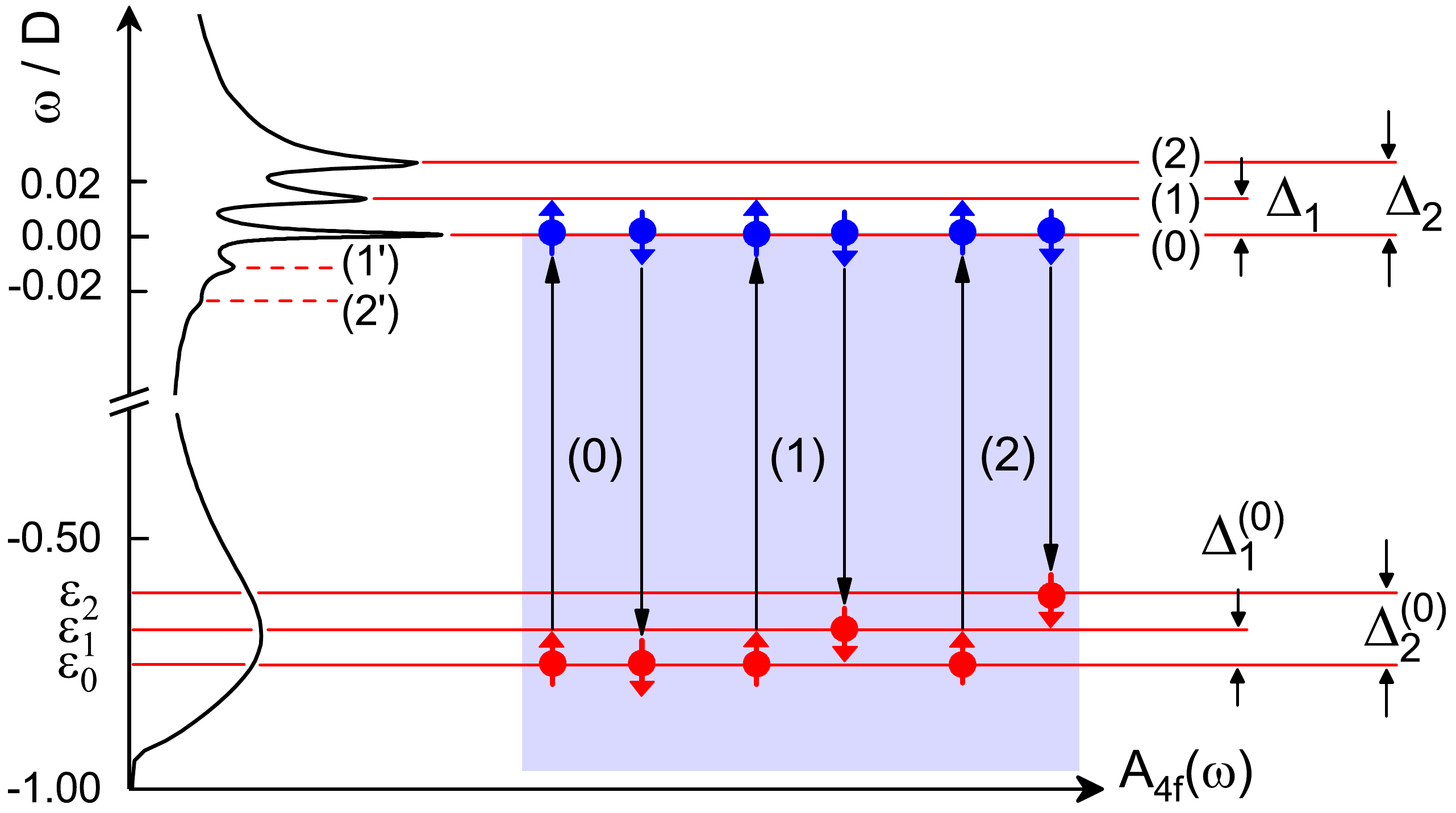}
\caption{Ce $4f$ spectral density for $T\ll T_K^*<\Delta_1$ 
(black curve) as calculated by DMFT for CeCu$_6$. 
The long arrows represent the hybridization processes 
between Ce $4f$ electrons (red) and conduction electrons (blue dots and shaded 
region) generating the Kondo and CEF resonances.
The energy scale around the Fermi energy 
($\omega\hspace*{-0.05cm}=\hspace*{-0.05cm}0$) is stretched by a factor 5.} 
\label{fig:Figure1}
\end{figure}

\textit{CEF resonances.}---CEF satellite structures in heavy-fermion systems have been previously
observed by photoemission \cite{Reinert01,Ehm07,Klein08} and scanning
tunneling \cite{Wirth11,Haze19} spectroscopy. 
In order to understand their impact,
one must realize that CEF resonances originate from the same strong 
correlation effect that generates the low-energy Kondo resonance 
itself \cite{Kroha03,Ehm07}. 
In the orthorhombic lattice structure of CeCu$_{6-x}$Au$_x$, the $j=5/2$
ground-state multiplet of the Ce $4f$ orbitals is split by the CEF into three
Kramers doublets, denoted by $\varepsilon_0$, $\varepsilon_1$, $\varepsilon_2$
(see Fig.~\ref{fig:Figure1}). Only one of the CEF states is 
significantly occupied at any time due to
strong Coulomb repulsion within the Ce $4f$ orbitals. By the hybridization of
these orbitals with the conduction-electron states, a Ce $4f$ electron can
fluctuate from the ground-state Kramers doublet $\varepsilon_0$ into a
conduction state near $\varepsilon_F$ and back to $\varepsilon_0$, shown as
process (0) in Fig.~\ref{fig:Figure1}. This process involves spin exchange
with the conduction electrons and is quasielastic (final and initial energies
are equal, $\varepsilon_0$), i.e., in resonance with the low-energy conduction
electron states \cite{Schrieffer66,Zamani16}. The singular quantum spin-flip scattering thus generates the narrow Kondo resonance in the Ce $4f$ spectrum at $\varepsilon_F$, shown as peak (0) in Fig.~\ref{fig:Figure1}. Alternatively, the $4f$ electron can end up in one of the CEF-excited levels, $\varepsilon_m$ ($m=1,\,2$) instead of $\varepsilon_0$ [process (1) in Fig.~\ref{fig:Figure1}]. Involving again singular quantum spin-flip transitions, this process generates another narrow resonance, albeit shifted in energy by the final-state excitation energy $\Delta_m=\varepsilon_{m}-\varepsilon_0+\delta\Delta_m$, i.e., by the bare CEF excitation energy $\Delta_m^{(0)}=\varepsilon_m-\varepsilon_0$ and additional many-body renormalizations $\delta\Delta_m$ [peak (1) in Fig~\ref{fig:Figure1}]. For each of the CEF satellites (1), (2) there exists a mirror satellite (1'), (2') 
\cite{Reinert01,Ehm07,Kroha03} shifted downward by $-\Delta_m$,
see Fig.~\ref{fig:Figure1}. The mirror satellites appear as weak peaks or
shoulders only, because they originate from transitions 
$\varepsilon_{1,2}\to\varepsilon_0$ whose initial states $\varepsilon_{1,2}$
have only a small, albeit nonzero (due to hybridization with the 
conduction band) virtual population even at $T\ll\Delta_1$.
From the above discussion, it is clear that the CEF satellite resonances are of 
spin-scattering origin, just like the Kondo resonance itself. Thus, for 
$k_BT_{K,m}\lesssim k_BT\ll\Delta_1$ their weight has a logarithmic
temperature dependence, and their width is renormalized by many-body effects
to exponentially small values of 
$T_{K,m}\approx D\,\exp\left[-{1}/{(2N_FJ_m)}\right]$
\cite{Hewson93,Kroha03},
where $D$ is the conduction half bandwidth, $N_F$ the (unrenormalized) density of states at the Fermi level, and $J_m$ the effective spin exchange coupling of the conduction electrons with the CEF level $m=0,\,1,\,2$, up to higher-order
renormalizations \cite{Kroha03,Ehm07}. In fact, this narrow width makes the CEF satellite resonances energetically separated and resolvable in spectroscopic experiments \cite{Reinert01,Ehm07,Wirth11} for $k_BT\ll\Delta_1$, while the hybridization width of the single-particle levels $\varepsilon_0$, $\varepsilon_1$, $\varepsilon_2$ is orders of magnitude larger than their splitting $\Delta_m^{(0)}$, see Fig.~\ref{fig:Figure1}. As the temperature is raised to $k_BT\approx \Delta_1$, at least one of the CEF-excited satellites becomes thermally occupied and acts as effectively degenerate levels, leading to the effective high-temperature Kondo scale 
$T_K^{\mathrm{(high)}}\approx D\,\exp\left[-{1}/{(2N_F\sum_m' J_m)}\right]$,
where the sum $\sum_m'$ runs over the CEF levels $m$ with significant thermal occupation. 

To quantify this behavior, we performed DMFT calculations for the multiorbital Anderson lattice model with three local levels $\varepsilon_m$,
corresponding to the three Kramers doublets of the $j=5/2$ Ce $4f$ ground-state
multiplet in CeCu$_6$. The near-single occupancy of the Ce $4f$ shell was 
enforced by a strong interlevel repulsion, $U\to\infty$. 
We chose the bare model parameters such that the DMFT produces the values for $T_K^*$ and the CEF splittings reported in the literature \cite{Stroka93,Goremychkin93,Witte07,Klein08} (see \cite{supplement} for details). The resulting DMFT spectra $A_{4f}(\omega)$ in Fig.~\ref{fig:Figure3}\,(a) exhibit the crossover from the high-temperature scale $T_K^{\mathrm{(high)}}\approx 200~\mathrm{K}~\widehat{=}~17$~meV to the low-energy Kondo scale $T_K^*=T_{K,0}\approx 6~\mathrm{K}~\widehat{=}~0.52$~meV.

\textit{Time-resolved terahertz spectroscopy.}---We 
investigate the temperature dependence of the QP spectral weight in the
CeCu$_{6-x}$Au$_{x}$ system. CeCu$_{6}$ has a Kondo lattice
scale of ${T_K^*} \approx 6$\,K and CEF excitations at $\Delta_1=7$\,meV and
$\Delta_2=13$\,meV \cite{Stroka93,Goremychkin93,Witte07}. CeCu$_{6-x}$Au$_{x}$
undergoes a magnetic QPT at $x=0.1$
\cite{Loehneysen07}. 
We generate terahertz pulses of 1.5 cycles duration, covering a
frequency range of 0.1--3\,THz, by optical rectification in a 0.5\,mm
(110)-oriented ZnTe crystal. We radiate these pulses onto samples of the
Fermi-liquid compound CeCu$_6$ and of the quantum-critical compound
CeCu$_{5.9}$Au$_{0.1}$ (see \cite{supplement} for details). 
The terahertz radiation induces symmetry-allowed dipole transitions from the
CEF-split heavy-fermion bands to the light band of Ce5$d$--Cu4$s$ character.
The reflected terahertz electric field is detected via free-space electro-optic 
sampling on a 0.5\,mm
(110)-oriented ZnTe crystal that is optically bonded to a 2\,mm (100)-oriented
ZnTe crystal. In this way, time traces of the reflected signal are taken from $t = -4$\,ps to $+8.5$\,ps in steps of 0.04\,ps. All time traces are normalized by a factor such that the integrated intensity equals one, representing identical total reflected power. It has been shown previously \cite{Wetli18} that a correlated many-body state manifests itself in the reflected terahertz electric field as a temporally confined and delayed pulse whose delay time resembles the QP lifetime (inverse spectral width), its integrated weight the QP weight. In particular, for a band of heavy Kondo QPs with spectral width $k_BT_K^*$ the delay time is $\tau_K^*=h/k_BT_K^*$, with $h$ the Planck constant and $k_B$ the Boltzmann constant \cite{Wetli18}. The temporally delayed pulse thus provides a direct and background-free probe for the QP dynamics of strongly correlated states.

\begin{figure}[t!]
\includegraphics[width=0.98\columnwidth]{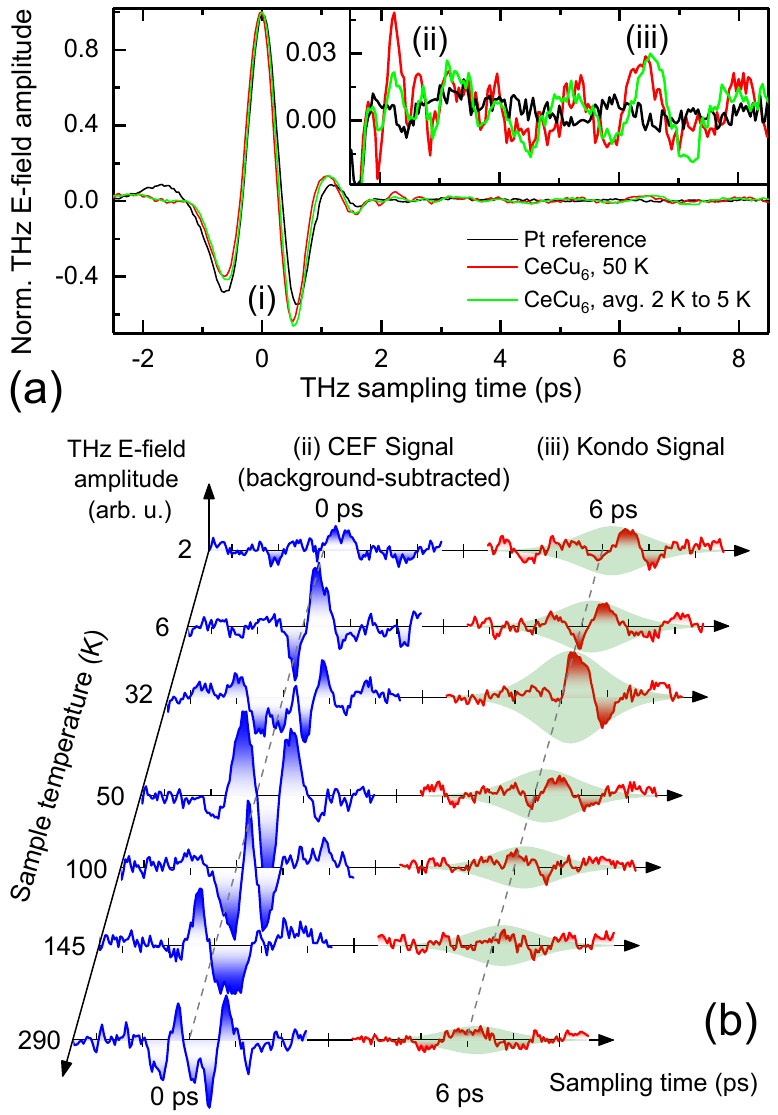}
\caption{(a) Time traces of the terahertz electric field reflected from a CeCu$_6$ sample at $T=50$\,K (red) in comparison to a Pt reference at $T=2.0$\,K (black) and the average of the CeCu$_6$ time traces for temperatures between 2.0 and 5.0 K (green). (b) Evolution of the background-subtracted CEF signal ($-2.5$ to $+2.5$\,ps) and the Kondo signal ($+3.5$ to $+8.5$\,ps) as the temperature decreases from 290 to 2\,K. The green-shaded region depicting the envelope of the Kondo signal is a solution of the nonlinear rate equation of Ref.~\cite{Wetli18} describing the relaxation of the terahertz-excited heavy-fermion system with a single local orbital.
Trivial delayed reflexes originating from the terahertz
generation crystal or the cryostat windows have been identified at times
$t>10$\,ps \cite{Wetli18}. 
}
\label{fig:Figure2}
\end{figure}

\begin{figure}[b!]
\includegraphics[width=0.96\columnwidth]{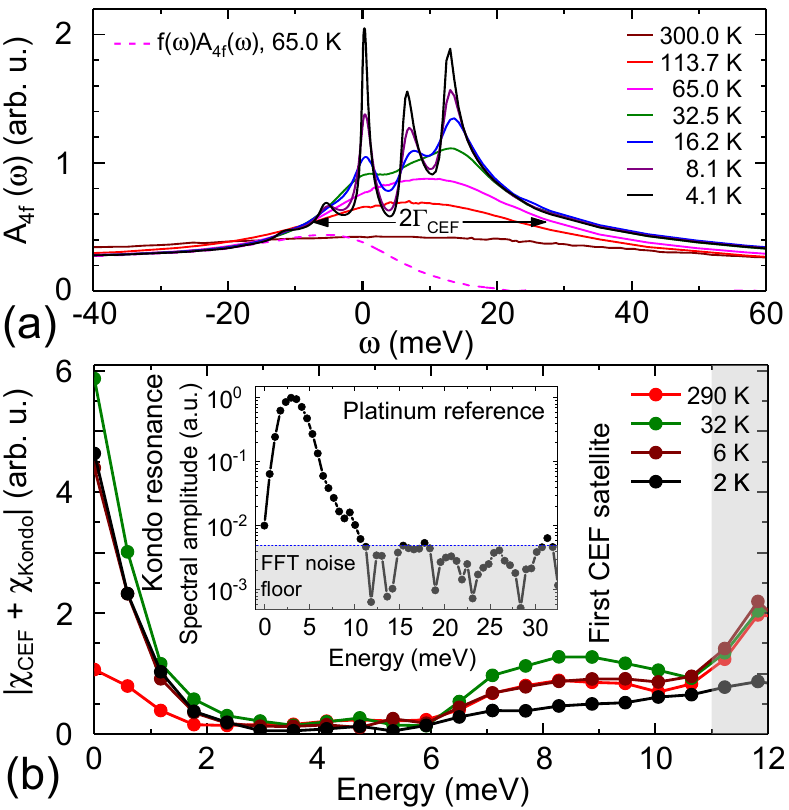}
\caption{(a) Temperature dependence of the momentum-integrated Ce $4f$ spectral density in CeCu$_6$ (solid lines), as calculated by DMFT for the Anderson lattice model. The occupied spectral density at 65.0 K, $f(\omega)\,A_{4f}(\omega)$ (dashed line; $f(\omega)$ is the Fermi-Dirac distribution), visualizes that 
the spectral width, $\Gamma_{\mathrm{CEF}}=k_BT_K^{\mathrm{(high)}}\approx
17~\mathrm{meV}~\widehat{=}\,200$\,K, is not accounted for by thermal
broadening at $65.0$~K alone. (b) Magnitude spectrum of the CEF and Kondo
responses of CeCu$_6$. The Kondo response peaks near zero energy, and has a
width of 0.56\,meV ($\widehat{=}\,6.7$\,K) which agrees very well with
$k_BT_K^*$. The first CEF satellite is seen at 8\,meV,
close to the literature value \cite{Stroka93,Goremychkin93,Witte07}.
The shaded region is beyond the spectral width of 
the terahertz excitation, i.e., governed by noise. 
The inset shows the Pt reference, equivalent to the incident terahertz spectrum
$|E_{\mathrm{in}}(\omega)|$.}
\label{fig:Figure3}
\end{figure}

\begin{figure}[t]
\includegraphics[width=0.96\columnwidth]{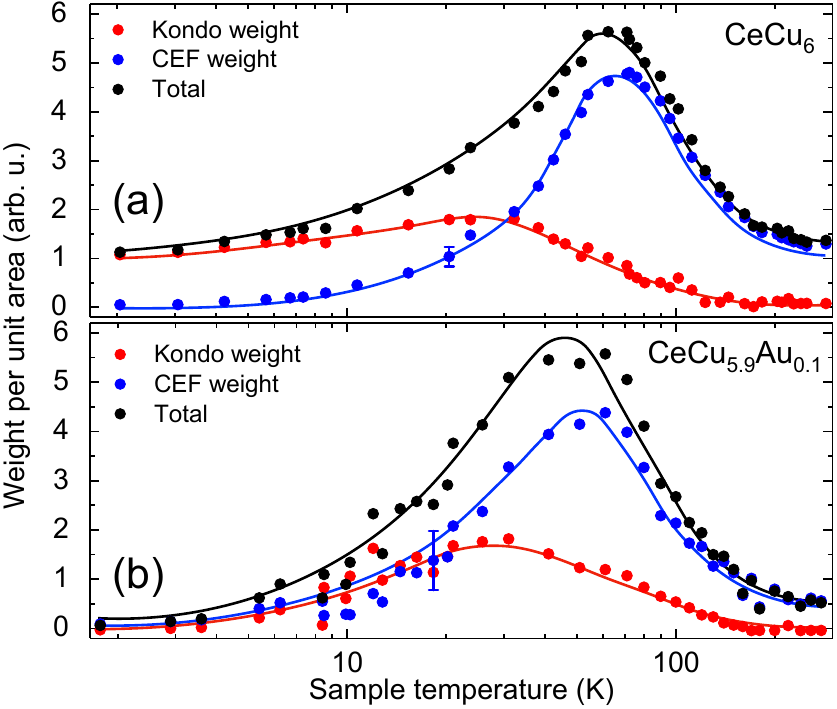}
\caption{Temperature dependence of the occupied weights of the Kondo (red) and
  of the CEF (blue) bands as well as the sum of the two (black) for (a)
  CeCu$_6$ and (b) CeCu$_{5.9}$Au$_{0.1}$. The error bars result from 
averaging over ten measurements for each data point. 
}
\label{fig:Figure4}
\end{figure}

\textit{Analysis of spectral features.}---A terahertz 
signal reflected from a CeCu$_6$ sample is shown in
Fig.~\ref{fig:Figure2}\,(a) in comparison to a platinum reference signal.
It exhibits three distinct features labeled (i), (ii), (iii).
We note that other features visible are statistical fluctuations:
They show no systematic temperature dependence and average out over  
ten measurements.
In the time interval [$-2.5$\,ps, $+2.5$\,ps] 
the signal consists of two overlapping features (i) and (ii). 
The strong, instantaneous pulse (i) centered at $t = 0$\,ps appears almost 
identically in CeCu$_6$ and in Pt and is temperature independent
[not shown in Fig.~\ref{fig:Figure2}\,(a)]. 
It is the stimulated single-particle response of the light 
conduction electrons. In addition, there is 
a weaker feature (ii) visible as the wiggles superimposed on the 
wing of pulse (i). We observe that this signal 
does not appear in the Pt reference, has a reproducible, 
nonmonotonic temperature dependence (analyzed below), and vanishes 
in all measured time traces below $T=5$\,K. 
We use this temperature dependence to separate 
signal (ii) from the single-particle reflex (i):  We take the 
temperature average of the time traces taken between 2 and 5\,K 
and subtract it from each time trace within the time 
interval [$-2.5$\,ps, $+2.5$\,ps]. 
Finally, the pulse (iii), centered around 6~ps, has been identified 
earlier \cite{Wetli18} with the Kondo resonance by its characteristic 
temperature dependence [c.f. Fig.~\ref{fig:Figure2}(b)] 
and by its delay time agreeing well with the 
Kondo QP lifetime $\tau_K^*$. For the detailed analysis of this signal, 
see Ref.~\cite{Wetli18}. 

The resulting, correlation-induced and background-subtracted time traces 
are shown in Fig.~\ref{fig:Figure2}\,(b). 
The amplitude of the signal (iii) from the 
heavy Kondo band rises with decreasing temperature and remains finite 
at the lowest temperatures, with some decrease due to the 
vicinity of the QPT \cite{Wetli18}. The background subtraction reveals 
that the reflex (ii) is not only located around $2$\,ps, but also extends
over a wider time range centered around the short delay time of 
$\tau_{\mathrm{CEF}}\approx 0.25$\,ps. It also rises with decreasing
temperature, but reaches a maximum near $T=60$\,K and then decreases to
undetectably small values for $T\lesssim 5$\,K. Such nonmonotonic behavior is
a clear signature of the CEF resonances, as can be seen from the DMFT
calculations of Fig.~\ref{fig:Figure3}\,(a): On reducing the temperature, 
a single, broad resonance of
width $\Gamma_{\mathrm{CEF}}=k_BT_K^{\mathrm{(high)}}\approx 17$\,meV 
rises up, corresponding to
the increase of the signal (ii) in Fig.~\ref{fig:Figure2}\,(b). 
Below about $60$\,K the broad resonance in
Fig.~\ref{fig:Figure3}\,(a) splits into three individual, sharp CEF peaks of
which only the lowest one is occupied towards $T=0$. Correspondingly, 
the short-delayed signal (ii) in Fig.~\ref{fig:Figure2}\,(b) 
disappears, and part of its weight reappears in the Kondo pulse (iii) with
$\tau_K^*=h/k_BT_{K,0}=h/k_BT_K^*=6$\,ps delay time. Hence, we have
unambiguously associated the time traces (ii) in Fig.~\ref{fig:Figure2}\,(b)
to the broad CEF resonance of Fig.~\ref{fig:Figure3}\,(a).

Further evidence for the CEF satellites is provided from the spectral analysis
using the (nonequilibrium) response 
function $\chi(\omega)=E_{\mathrm{out}}(\omega)/E_{\mathrm{in}}(\omega)$,
defined as the ratio of the Fourier transforms of the reflected 
signal $E_{\mathrm{out}}(\omega)$ and the incident light 
pulse $E_{\mathrm{in}}(\omega)$ (given by the Pt reference in the present
case). The magnitude spectrum of the sum of the response functions
$\chi_{\mathrm{CEF}}(\omega)$ and $\chi_{\mathrm{Kondo}}(\omega)$ resulting
from the traces (ii) and (iii) of Fig.~\ref{fig:Figure2}\,(b), respectively,
is shown in Fig.~\ref{fig:Figure3}\,(b). It clearly exhibits the Kondo
resonance and the first CEF satellite resonance, including their
characteristic temperature dependences.

\begin{figure}[b!]
\includegraphics[width=0.96\columnwidth]{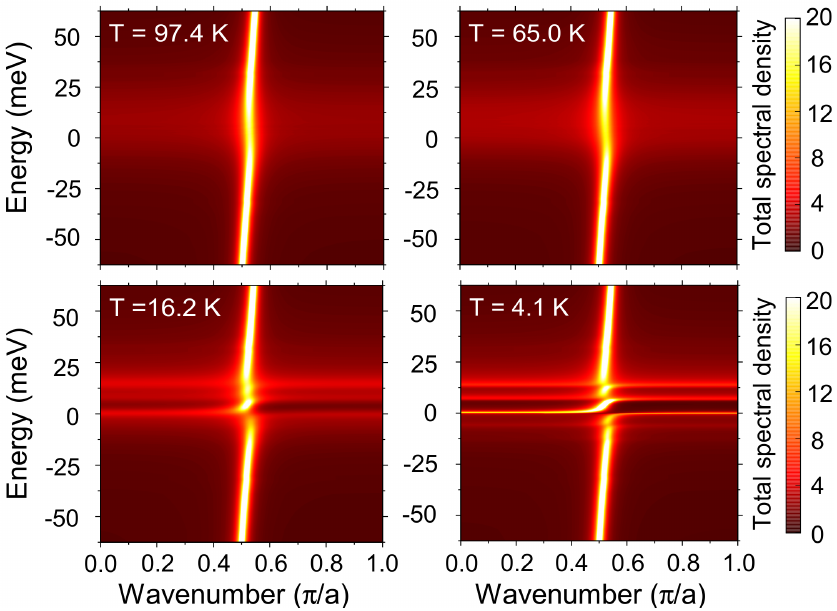}
\caption{DMFT band structure near $\varepsilon_F$ 
  of a three-orbital Anderson lattice for several
  temperatures. The model parameter values are chosen in order to resemble
  CeCu$_6$, $T_K^*\approx 6$\,K and $\Delta_1=7$\,meV, $\Delta_2=13$\,meV. The
  nearly vertical line represents the light conduction band. At low
  temperatures, $T\lesssim T_K^*$, the sharp, heavy Kondo and CEF satellite
  bands are well resolved, each one hybridizing with the conduction band. At
  high temperatures, $T\gtrsim \Delta_1$, they merge to a single, heavy band
  of shorter lifetime, but still of significant spectral weight at $\varepsilon_F$, thus maintaining a large Fermi volume up to $T\gtrsim T_K^{\mathrm{(high)}}\gg T_K^*$.}
\label{fig:Figure5}
\end{figure}

\textit{Large Fermi volume at high temperatures.}---We can now 
separately analyze  the Fermi volume change induced
by the low-temperature Kondo effect and by the CEF excitations. We integrate,
after background subtraction, the time traces of the squared terahertz electric
field over the time intervals [$-2.5$\,ps, $+2.5$\,ps] (ii) and [$+3.5$\,ps,
  $+8.5$\,ps] (iii) in Fig.~\ref{fig:Figure2}\,(b). The weights calculated in
this way represent directly the total occupation numbers of the heavy Kondo
and CEF satellite bands near $\varepsilon_F$, respectively, since the terahertz
excitation is sensitive to occupied states only. Thus, these
weights directly account for the correlation-induced Fermi volume change
\cite{Remark1,Feng18}. 
The temperature-dependent results are shown for CeCu$_6$ in
Fig.~\ref{fig:Figure4}\,(a). The Kondo as well as the CEF weights rise
logarithmically with decreasing temperature, confirming their Kondo-like 
origin. The Kondo weight reaches a maximum near $T=30$~K and
settles to a finite value at the lowest temperatures. The CEF weight dominates
the high-temperature behavior up to a characteristic temperature of
$T_K^{\mathrm{(high)}}\approx 200~\mathrm{K}~\widehat{=}~17~\mathrm{meV}$, read off
from the onset of its logarithmic rise 
and in good quantitative agreement with the DMFT result for the 
CEF resonance width at high temperature [Fig.~\ref{fig:Figure3}\,(a)]. 
The CEF weight vanishes gradually below $\sim 60$\,K, 
when the CEF satellite occupation gets frozen out and is transferred 
to the rising Kondo weight.  
The smooth crossover is confirmed by the momentum-resolved 
DMFT calculations, shown in Fig.~\ref{fig:Figure5}. 
It is in line with ARPES 
results on other heavy-fermion compounds \cite{Feng17,Allen17,Generalov18}.
Thus, our experimental findings and theoretical
calculations reveal consistently that an enlarged Fermi volume persists at
temperatures much higher than $T_K^*$ due to the CEF satellite contributions,
but it is carried by the Kondo spectral weight alone when the temperature is
lowered below the CEF splitting $\Delta_1$. The results of analogous
measurements for the quantum-critical compound CeCu$_{5.9}$Au$_{0.1}$ are
shown in Fig.~\ref{fig:Figure4}\,(b). We see that the CEF contribution to the
Fermi volume is almost identical to that in the Fermi-liquid compound
CeCu$_{6}$ for all temperatures. However, the Kondo spectral weight is seen to
vanish at the quantum-critical point in CeCu$_{6-x}$Au$_{x}$.

\textit{Conclusion.}---Time-domain THz spectroscopy provides a nearly 
background-free probe
of quasiparticle dynamics in correlated electron systems with a characteristic
energy scale in the terahertz range.
Our measurements and DMFT calculations on CeCu$_{6-x}$Au$_x$ 
show that the spectral weight near the Fermi level is governed by a 
heavy-quasiparticle state whose width crosses over from the CEF-induced 
high-energy scale $T_K^{\mathrm{(high)}}$ to the Kondo lattice scale $T_K^*$  
at low temperature. Terahertz reflection acts as a ``time filter'' separating 
the CEF excitations from the Kondo resonance by different reflex 
delay times. Employing this, we showed that at high temperatures 
$T\approx T_K^{\mathrm{(high)}}\gg T_K^*$ the Fermi volume is enlarged
by the CEF excitations, both in the Fermi liquid phase (CeCu$_{6}$) and
in the quantum-critical compound CeCu$_{5.9}$Au$_{0.1}$. At low temperatures 
$T<\Delta_1$, the large Fermi volume is carried by the 
ground-state Kondo band in CeCu$_{6}$, but collapses at the QPT
\cite{Wetli18}. 
This reconciles, within a single experiment, the existence of a large Fermi
volume at $T\gg T_K^*$ with vanishing Kondo weight at the QPT.
Since this CEF-induced mechanism appears to be 
generic, we expect similar behavior of other Ce- and Yb-based 
heavy-fermion compounds \cite{Kummer15,Feng17,Allen17,Generalov18} 
in forthcoming terahertz experiments.

\begin{acknowledgments}
The authors gratefully acknowledge insightful discussions with S.~Kirchner,
K.~Kliemt, C.~Krellner, K.~Matho, S.~Wirth, and G.~Zwicknagl. 
M. F. thanks CEMS at RIKEN for support of his research sabbatical.
This work was financially supported by the Swiss National Science 
Foundation (SNSF) 
via Projects No.~200021\underbar{\ }147080 (M.F., C.W.) and 
200021\underbar{\ }178825 (M.F., S.P.)
and by the Deutsche Forschungsgemeinschaft (DFG) via Grant 
No.~SFB/TR 185-C4 (J.K., F.Z.).
\end{acknowledgments}

\vfill

\onecolumngrid

\newpage

\begin{centering}

{\large{\bf Supplemental Material for \\[0.2cm]
Fermi volume evolution and crystal field excitations 
in heavy-fermion compounds probed by time-domain 
terahertz spectroscopy}}\\[0.4cm]

S. Pal$^{1}$, C. Wetli$^{1}$, F. Zamani$^{2}$, O. Stockert$^{3}$, 
H. v. L\"ohneysen$^{4}$, M. Fiebig$^{1}$, J. Kroha$^{2,5}$\\[0.4cm]

{\small
$^{1}$ {\it Department of Materials, ETH Z\"urich, 8093 Z\"urich, Switzerland}\\
$^{2}$ {\it Physikalisches Institut and Bethe Center for Theoretical Physics,
Universit\"at Bonn, Nussallee 12, 53115 Bonn, Germany}\\
$^{3}$ {\it Max Planck Institute for Chemical Physics of Solids, 01187 Dresden, Germany}\\
$^{4}$ {\it Institut f\"ur Festk\"orperphysik and Physikalisches Institut, Karlsruhe Institute of Technology, 76021 Karlsruhe}\\ 
$^{5}$ {\it Center for Correlated Matter, Zhejiang University, 
Hangzhou, Zhejiang 310058, China}}\\[0.6cm]

This supplement describes specifications of the experimental setup 
as well as the definition of the multi-orbital Anderson lattice model
and details of the dynamical mean-field theory (DMFT) calculations with 
the non-crossing approximation (NCA) as the impurity solver. \\ 

\vspace*{0.8cm}

\end{centering}

\setcounter{equation}{0}
\setcounter{figure}{0}
\renewcommand{\theequation}{S\arabic{equation}}
\renewcommand{\thefigure}{S\arabic{figure}}

\twocolumngrid

\subsection{Terahertz time domain reflection spectroscopy}
\begin{figure}[b]
\includegraphics[width=0.94\columnwidth]{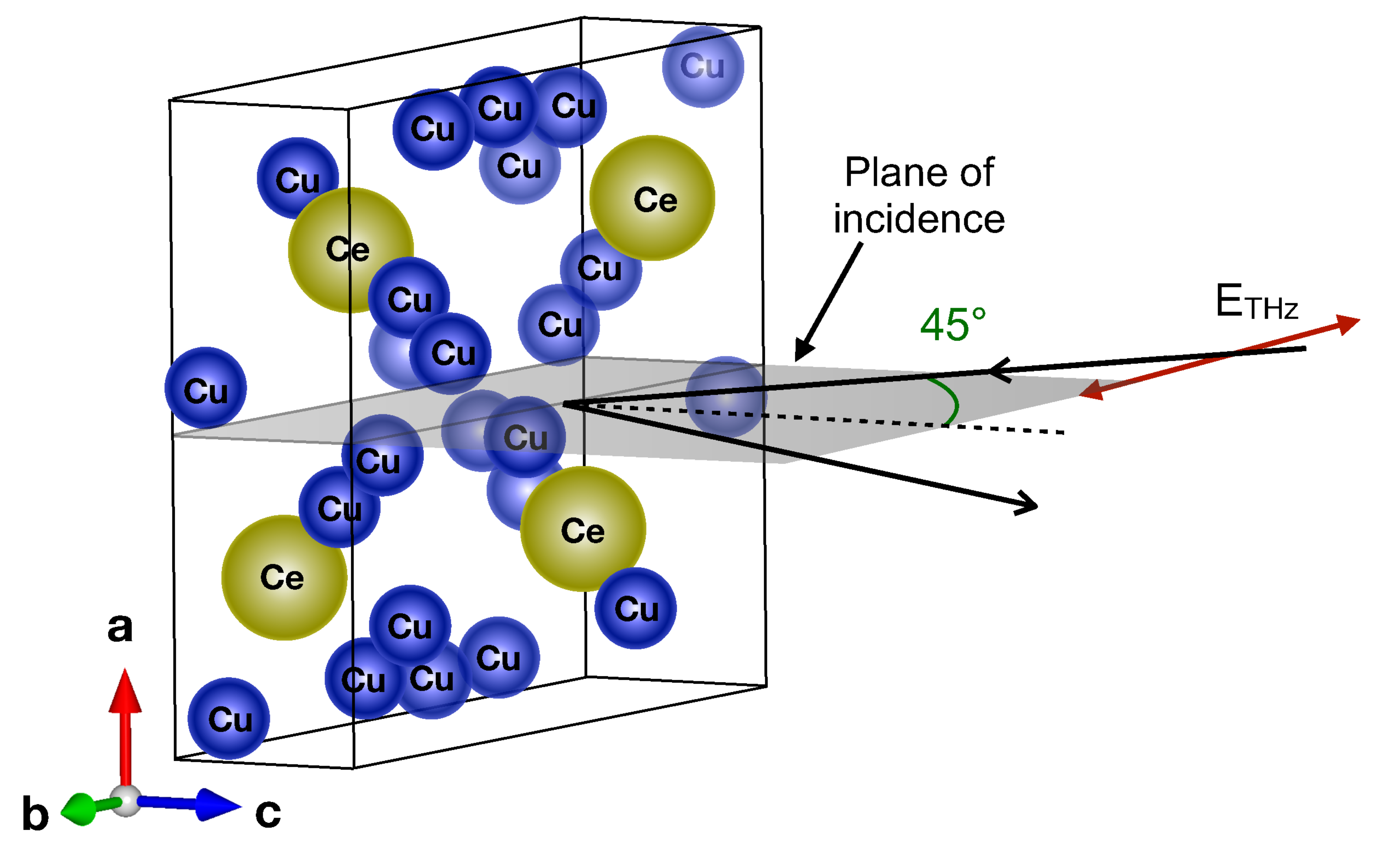}
\caption{
The crystallographic axes of the sample and experimental geometry 
using P-polarized light, with a 45$^o$ incidence angle.
} 
\label{fig:FigureS1}
\end{figure}

\textit{Geometry of experimental setup.} The terahertz time domain spectroscopy is performed in reflection geometry. All CeCu$_{6-x}$Au$_{x}$ samples were cut from single crystals of orthorhombic crystal structure, and faces perpendicular to the crystallographic c-axis were polished using SiC. The samples were then mounted in a temperature-controlled Janis SVT-400 helium reservoir cryostat such that the c axis is parallel to the plane of the optical table. Linearly polarized terahertz radiation with a spectral range of 0.1--3~THz was incident on the sample at an angle of 45$^{o}$, the terahertz electric field $E_{\mathrm{THz}}$ oriented perpendicular to the crystallographic  a axis, see Fig.~\ref{fig:FigureS1}. Hence, $E_{\mathrm{THz}}$ is P-polarized, i.e., parallel to the plane of incidence.

\textit{THz pulse generation and analysis.} Single-cycle terahertz pulses of a few nanojoules are generated by optical rectification in a 0.5 mm ZnTe-(110) single crystal, using 90~\% of a Ti:Sapphire laser output (wavelenth 800 nm, pulse duration 120 fs, pulse repetition rate 1 kHz, 2 mJ per pulse). The remaining 10~\% of the fundamental 800 nm laser pulse are used as a probe (or gating) pulse for the electrooptic sampling of the reflected terahertz wave. The terahertz and the gating beams are collinearly focused onto a ZnTe-(110) detection crystal. The terahertz-induced ellipticity of the probe light is measured using a quarter-wave plate, a Wollaston polarizer and a balanced photodiode (BPD). The signal from the BPD is then analyzed with a lock-in amplifier. In order to increase the accessible time delay between the terahertz and the probe pulses, Fabry-P\'erot resonances from the faces of the 0.5 mm thick ZnTe-(110) crystal are suppressed by extending the detection-crystal with a 2-mm-thick, terahertz-inactive ZnTe-(100) crystal that is optically bonded to the back of the detection crystal.

\vspace*{-0.2cm}

\subsection{Multi-orbital Anderson lattice model}
In the orthorhomic lattice structure of CeCu$_{6-x}$Au$_{x}$, the fourteen
(including spin) orbitals of the Ce 4f shell are split by spin-orbit 
coupling into a total angular momentum $j=5/2$ ground-state multiplet and 
a $j=7/2$ excited multiplet, with a spin orbit splitting of 
$\Delta_{SO}\approx 250$~meV \cite{S.Ehm07}. The ground-state multiplet is 
further split by the crystal electric field (CEF) into three Kramers 
doublets with excitation energies 
$\Delta_{m}^{(0)}=\varepsilon_{m}-\varepsilon_{0}$, $m=1,\,2$, 
above the ground-state energy $\varepsilon_0$, see Fig.~1 of the main article.
The CeCu$_{6-x}$Au$_{x}$ system with the three low-lying Ce 4f Kramers 
doublets accessible by thermal or terahertz excitation is, thus, described 
by the Anderson lattice model with three local orbitals,
\begin{eqnarray}
H&=&
\sum_{\bm{k},\sigma}\varepsilon_{\bm{k}}\,
c^{\dagger}_{\bm{k}\sigma}c^{\phantom{\dagger}}_{\bm{k}\sigma}+
\sum_{i,m,\sigma}\,\varepsilon_{m}
d^{\dagger}_{im\sigma}d^{\phantom{\dagger}}_{im\sigma} \nonumber \\
&+&\frac{U}{2} \sum_{i,\,(m,\sigma)\neq (m'\hspace*{-0.05cm},\sigma')}
d^{\dagger}_{im\sigma}d^{\phantom{\dagger}}_{im\sigma}\,d^{\dagger}_{im'\sigma'}d^{\phantom{\dagger}}_{im'\sigma'} \\
&+&\sum_{\bm{k},i,m,\sigma} \left(
V_{\bm{k}m}\, \mathrm{e}^{-\mathrm{i}\bm{kr}_i}\,
c^{\dagger}_{\bm{k}\sigma}d^{\phantom{\dagger}}_{im\sigma}+
V_{m\bm{k}}^*\, \mathrm{e}^{\mathrm{i}\bm{kr}_i}\,
d^{\dagger}_{im\sigma}c^{\phantom{\dagger}}_{\bm{k}\sigma} 
\right)\, ,
\nonumber
\end{eqnarray}
where $c^{\dagger}_{\bm{k}\sigma}$, $c^{\phantom{\dagger}}_{\bm{k}\sigma}$
are the field operators for conduction electrons in momentum and 
spin state $|\bm{k},\sigma\rangle$ with dispersion $\varepsilon_{\bm{k}}$. 
$d^{\dagger}_{im\sigma}$, $d^{\phantom{\dagger}}_{im\sigma}$ are the operators 
for electrons in the Ce 4f Kramers doublet states $|m,\sigma\rangle$
located at the lattice sites $i$,
with single-particle binding energies $\varepsilon_m$,
$m=0,\,1,\,2$, $\sigma=\pm 1/2$.
The third term of the Hamiltonian represents the strong Coulomb repulsion $U$
between electrons in any of the Ce 4f orbitals which effectively 
enforces single occupance of the Ce 4f shell in accordance with 
the valence of the cerium atoms in CeCu$_{6-x}$Au$_{x}$. For the 
calculations we, therefore, take $U\to\infty$.  The fourth term
describes the hybridization between the conduction electron states and the 
Ce 4f states $|m,\sigma\rangle$ on each lattice site $i$.
We apply standard dynamical mean-field theory (DMFT) to compute the 
energy- and momentum-dependent spectral functions of the hybridizing
(light) conduction band and the three CEF-split 4f bands  
of this interacting system, shown in Fig.~5 of the main article.  
As the DMFT impurity solver we use the multi-orbital generalization 
of the slave-boson (SB) representation within the  
non-crossing approximation (NCA) \cite{S.Ehm07,S.Costi96}, 
because it is able to 
describe the spectra over the complete width of the uncorrelated 
conduction band of $O(10~\mathrm{eV})$,
while maintaining a resolution of better than $O(0.1~\mathrm{meV})$
near the Fermil level, necessary to resolve all CEF resonances at low 
temperatures $T<T_K^*$. 
For a single Anderson impurity in the $U\to\infty$ limit, the NCA is known 
to correctly describe the width and temperature dependence of the 
Kondo and CEF spectral features from high $T$ down to well below $T_K^*$ 
\cite{S.Ehm07,S.Costi96}.

In the SB formulation, each quantum state $|m,\sigma\rangle$ 
of the Ce 4f shell on site $i$ is represented by a fermionic 
field $f^{\dagger}_{im\sigma}$ and the unoccupied Ce 4f shell by a 
bosonic field $b^{\dagger}_{i}$, such that the electron operators read
$d^{\dagger}_{im\sigma} = f^{\dagger}_{im\sigma}b_{i}$, with the local 
operator constraint 
$\sum_{m\sigma} f^{\dagger}_{im\sigma}f^{\phantom{\dagger}}_{im\sigma}+
b^{\dagger}_{i}b^{\phantom{\dagger}}_{i} = 1$. 
The local (retarded) pseudofermion propagator 
$G_{f\,mm'\,\sigma}(\omega)$ and its selfenergy 
$\Sigma_{f\,mm'\,\sigma}(\omega)$ are nondiagonal matrices in Ce 4f orbital
space,
\begin{equation}
G_{f\,mm'\,\sigma}(\omega) =
\left[ (\omega -\lambda -\varepsilon_m)\delta_{mm'}
-\Sigma_{f\,mm'\,\sigma}(\omega)
\right]^{-1}\, ,
\label{Gf}
\end{equation}
and the (retarded) slave boson propagator reads,
\begin{equation}
G_{b}(\omega) =
\left( \omega -\lambda -\Sigma_{b}(\omega) \right)^{-1}\, .
\label{Gb}
\end{equation}
$\lambda$ is a chemical-potential parameter used to enforce the 
constraint exactly by the limit $\lambda\to\infty$ at the end of the 
calculation \cite{S.Costi96}. 
The multi-orbital NCA equations then read \cite{S.Kroha05},
\begin{eqnarray}
\hspace*{-0.4cm}&\Sigma&\hspace*{-0.0cm}_{f\,mm'\,\sigma}(\omega)=
             \Gamma _{mm'}\int 
              {d}\varepsilon\,
               f(\varepsilon )
              \tilde{A}_{c\sigma}(-\varepsilon)G_{b}(\omega +\varepsilon )
              \phantom{xxxxii}
              \label{sigfNCAm}\\
&\Sigma&\hspace*{-0.0cm}_{b}(\omega )
              =
              \sum _{mm '\sigma}\hspace*{-0.1cm}\Gamma_{m 'm }\hspace*{-0.1cm}
              \int 
              {d}\varepsilon\,
              f(\varepsilon )\tilde{A}_{c\sigma}(\varepsilon)
              G_{f\,mm\, '\sigma}(\omega +\varepsilon )\, .
              \label{sigbNCAm}
\end{eqnarray}
Here, $f(\varepsilon)$ is the Fermi distribution function, 
$\tilde{A}_{c\sigma}(\varepsilon)$ the 
momentum-integrated $c$-electron spectral density without scattering 
from the DMFT impurity site, and 
\begin{equation}
\Gamma_{m m' } = \sum_{\bm{k}} V^*_{m\bm{k}}\, |\mathrm{Im}\,
G^{(0)}_{c\,\bm{k}\sigma}(0)|\, V_{\bm{k}m'} 
\end{equation}
is the effective hybridization matrix, with 
$G^{(0)}_{c\,\bm{k}\sigma}(0)$ the bare $c$-electron propagator at the 
Fermi energy. The NCA equations (\ref{sigfNCAm}), (\ref{sigbNCAm}) 
together with Eqs.~(\ref{Gf}), (\ref{Gb}) are solved selfconsistently 
by iteration. See \cite{S.Costi96} for an efficient numerical treatment. 
Denoting the slave-boson and pseudo\-fermion spectral functions by
$A_b(\varepsilon)=-\mathrm{Im}\,G_{b}(\varepsilon)/\pi$ and
$A_{f\,mm'\,\sigma}(\varepsilon)=-\mathrm{Im}\,G_{f\,mm'\,\sigma}(\varepsilon)/\pi$,
respectively,
the Green's function for electrons in the Ce 4f orbitals is then 
obtained as ($\beta=1/(k_BT)$) 
\begin{eqnarray}
&G&\hspace*{-0.0cm}_{d\, mm\, '\sigma}(\omega )
         =
         \int  {d}\varepsilon\,  {\rm e}^{-\beta\varepsilon}
         [ G_{f\,mm'\,\sigma}(\omega +\varepsilon )A_{b}(\varepsilon )
          \nonumber\\
         &\ &\hspace*{3cm}-A_{f\,mm'\,\sigma}(\varepsilon )
                   G_{b}(\varepsilon -\omega ) ] \ ,
          \label{gdNCAm}
\end{eqnarray}
which is subsequently fed into the DMFT loop. 

For the numerical evaluations we considered, for simplicity, 
a three-dimensional, cubic tight-binding lattice. The parameter values
(in units of the bare conduction half bandwidth $D$, where for copper 
$D\approx 8$~eV) were adjusted as
\begin{eqnarray*}
\begin{matrix}
\varepsilon_0=-0.5, &\quad\varepsilon_1=-0.499, &\quad\varepsilon_2=-0.0498 \\
\Gamma_{00}=0.063, &\quad\Gamma_{11}=0.042, &\quad\Gamma_{22}=0.032, \\  
\Gamma_{01}=0.0053, &\quad\Gamma_{12}=0.0105, &\quad\Gamma_{02}=0.0053
\end{matrix} 
\end{eqnarray*}
These values reproduce the experimentally known values for CeCu$_6$ of
$T_K^*\approx 6$~K and CEF splittings $\Delta_1=7$~meV,
$\Delta_2=13$~meV \cite{S.Stroka93,S.Goremychkin93,S.Witte07}.


\begin{thebibliography}{[1]}
\bibitem{Remark1}
When the conduction band has holelike character (negative dispersion), 
its hybridization with the heavy Ce states leads to a 
decrease of the Fermi momentum, as discussed in \cite{Feng18} for 
the holelike compound CeRhIn$_5$. This is consistent with our 
present findings of a Kondo-induced Fermi momentum increase, since 
the hole number counts negative.
\end{thebibliography}
\end{document}